# Effective Listings of Function Stop words for Twitter

Murphy Choy

**Abstract**

Many words in documents recur very frequently but are essentially meaningless as they are used to join words together in a sentence. It is commonly understood that stop words do not contribute to the context or content of textual documents. Due to their high frequency of occurrence, their presence in text mining presents an obstacle to the understanding of the content in the documents. To eliminate the bias effects, most text mining software or approaches make use of stop words list to identify and remove those words. However, the development of such top words list is difficult and inconsistent between textual sources. This problem is further aggravated by sources such as Twitter which are highly repetitive or similar in nature. In this paper, we will be examining the original work using term frequency, inverse document frequency and term adjacency for developing a stop words list for the Twitter data source. We propose a new technique using combinatorial values as an alternative measure to effectively list out stop words.

**Introduction**

Text mining comprises of a series of tasks that includes selection of approach, parameter setting and the creation of a stop word list (Keogh et al., 2004; Xu and Wunsch, 2005). The creation of a stop word list is often viewed as an essential component of the text mining which requires manual labor and investigations to produce. Stop words lists are rarely investigated and validated compared to the results of the mining process or mining algorithm. The lack of research into stop words list creation resulted in extensive use of pre-existing stop word lists which might not be suitable given the differences in the context of the textual sources. Research in the area has identified the weaknesses of standardized stop words list (Chakrabarti et al., 1997; Chakrabarti et al., 1998; Silva and Ribeiro, 2003).

With the spread of social media platforms and adoption of such technologies in business and daily life, social media platforms have become one of the most important forms of communication for internet users and companies. Some companies are using facebook and twitter system to provide real time interaction with their customers. These social media platforms are beneficial to companies building consumer brand equity (Jothi, 2011). The platforms also act as low cost effective measures to manage complex relations between companies and consumers. The nature of social media also promotes open and transparent resolution of disputes and allows for greater visibility of the disputes to the senior management. Social Media has also proven to be very effective in communicating news such as the occurrence of earthquakes (Sazaki, Okazaki and Matsuo, 2010; Earle, 2010) and political office election (Tumasjan et. Al., 2010;Mungiu-Pippidi, 2009).

The enormous amount of textual information from twitter and social media requires extensive amount of data preparation and analysis to reap any benefits. There are many approaches to analyze the data. However, due to the nature and assumptions of the techniques as well as the huge amount of data collected, the data quality has to be of a very high level of quality in order to be effective (Cooley et al., 1999; Redman, 1998; Jung, 2004; Tayi and Ballou, 1998; Zhang et al., 2004). To improve the quality of textual data, many authors have proposed different techniques to extract an effective stop word list for a particular corpus (Salton, 1988; Rose et. al., 2010). In the next section, we will focus on the common approaches to the development of stop words list.

**Current approaches**

A stop words list refers a set of terms or words that have no inherent useful information. Stop words create problems in identification of key concepts and words from textual sources when they are not removed due to their overwhelming presence both in terms of frequency as well as occurrence in textual sources. Several authors (van Rijsbergen, 1979; Sinka and Corne, 2003). Manco et al. (2002) have argued for the removal of stop words which make the selection of the useful terms more efficient and reduce the complexity of the term structure. The current literature divides the stop words into explicit stop words and implicit stop words.

The common approach is to manually assemble a stop words list from a list of words. This approach is used by several authors(Van Rijsbergen, 1975; Fox, 1992) and has proven to be generally applicable to a variety of situation(Sinka and Corne, 2010). Even though the generic stop words lists generally achieved high accuracies and robust in nature, customized stop words lists occasionally outperforms especially in technical areas. These customized stop words lists were developed based on the entropy lists or unions of the standard stop lists with entropy lists mixed in (Silva and Ribeiro, 2003). Other authors held the opinion that any words that appear too rarely or were longer than a certain length should be removed (Koprinska et al., 2007).

There have been other attempts to use a variety of frequency measures such as term frequency, document frequency or inverse document frequency (Konchady, 2006; Manning and Schutze, 1999). Each of these measures has proven to be effective in extracting the most common words that appear in the documents. The combination of term frequency with inverse document frequency (tf-idf) measure was widely quoted by text books and papers (Konchady, 2006; Rose, et. Al., 2010) as the most popular implicit approach for creating a stop words list. In Rose et. Al. (2010), the authors proposed a new measure called the adjacency measure to establish whether a particular word is a stop word or a content word. In the next section, we will examine the algorithm described by Rose et. Al.

**Rapid Automatic Keyword Extraction Stop word list**

In the paper "Automatic keyword extraction from individual documents" by Rose et. Al., the authors describe a process to determine the usefulness of that word in describing the contents. Every word is identified and the word co-occurrences are calculated with a score is calculated for each word. Several scoring techniques based on the degree and frequencies of words were evaluated in the paper. In the paper, Adjacency frequency is defined as the number of times the word occurred adjacent to keywords. Keyword frequency is defined as the number of times the word occurred within keywords. The authors noted that selection by term frequency will increase the likelihood of content-bearing words to be added to the stop words list for a specialized topic that result in removal of critical information words. Rose et. Al. describes the adjacency algorithm as 'intuitive' for words that are adjacent to keywords are less likely to be useful than those that are in it. The authors subsequently tested the algorithm using several standardized documents and found the algorithm to be very effective.

However, there are several issues with the use of the adjacency measure.

1. Adjacency measure first assumes the presence of a keyword in which we can use to determine words that are adjacent. This results in the technique being usable only in the case where keywords are specified. In most textual sources, keywords are not available. In the case of twitter, while you can use query keywords, it may not be useful for general trend extraction from tweets.
2. Adjacent words might be descriptive words which cannot be found within the keywords. In this case, the measure punishes these words.
3. Adjacency measures assumes multiple keywords in order for the between keywords to be found. This is an unlikely situation given that keywords are likely to single words. This makes it very difficult to be applied to twitter or documents where the keywords are single words.

Given the restrictive nature of the RAKE stop words list generator, it is very difficult to apply the algorithm to a wide spectrum of text mining problems. In the next section, we will extend on the ideas given in Rose et. al.(2010) And present an effective algorithm in listing functional stop words using the combinatorial counts as measure of information value.

**Effective Listings of Functional Stop words using combinatorial counts**

The authors noted that while the adjacency-within factor cannot be easily computed, the combinatorial factor can be computed easily. The combinatorial factor is defined as the number of unique word combination that can be found in the collection of tweets given a start word. The mathematical form is expressed below.

$$TCF = \sum_{i=1}^{n} f(w_{p,n}, w_{p+1,n})$$

Where *n* is the number of tweets, *p* is the position of the word and $w_p$ is the word in the position p. The function f is the indicator function with the following behavior.

$$f = \begin{cases} 1, where\, w_p = w \\ 0, where\, w_p \neq w \end{cases}$$

Where *w* is the word that is being investigated.

The measure is computationally simple and implementable in a variety of programming languages natively. The combinatorial nature of the measure may not be intuitive. Any words can be linked by a number of words in a language to form meaning combinations. Words designed to convey a precise meaning needs to be linked up in a particular combination for the correct meaning to be conveyed. However, words which are commonly used as bridges in sentences will naturally accumulate a large number of combinations in any collection of documents or tweets. If the collection contains a strong theme or event, the words related will have smaller combinations of words. Theoretically, if there are certain words which are important, the number of combinations should only be one. For example, in any discussion about Linear Algebra, many of the technical terms used will naturally have little variations such as 'Linear Models', 'Complement Set'. This is in contrast to words such as 'in the' and 'that is'.

This measure is an alternative approach to the classical techniques of term-frequency and inverse-document frequency. This approach measures the information value of the word not through the conventional Kullback – Leibler framework but through the combinatorial nature of words. As opposed to measuring the information value of words to establish the stop words, the technique focuses on the extreme number of combinations that most non-meaningful words display to establish stop words. Moreover, the use of combinations allow us to naturally manage both words with high and low occurring frequency which presents a problem for the classical framework of TF*IDF without using transformation.

**Experimental setup**

To validate the prowess of the measure, we conducted experiments with several techniques commonly used in development of stop word list. For all the experiments conducted, we have selected 9 3-days periods containing tweets with the key word search of 'Earthquake'. Each of this period starts 24 hours before the beginning of an earthquake and last till 48 hours after the occurrence of the earthquake. The reason for selecting 9 different periods and earthquakes is to ensure that the experiments will be as unbiased as possible. The use of query based tweets is to ensure that we have some form of central themes which provides some kind of comparison for the words which are not useful or meaningful. This two conditions enable us to assess the overall performance for the techniques tested effectively and unbiased.

The control factor for this experiment is the Fox's and Manu's stop word list. The choice of having two stop word lists is to double validate the techniques as both stop word lists are commonly used for text mining purposes. At the same time, both stop word lists have different words which can be useful as a further comparison between the efficacies of the techniques. All the words found in both stop word lists are determined to be stop words in the tweets through human examinations of the tweets using random samples of 1000 unique tweets from each period. For the classical techniques such as term frequency and inverse document frequency, we varied the cutoff thresholds before determining the optimal threshold by calculating the precision of the generated list with the stop list for different range of values. In total, we generated about 10 lists per technique.

Once we have generated the lists, we then compare the list across the different levels of threshold in increasing level of liberty in allowing the word to be considered stop word. Both precision and recall are calculated together with F-measure by comparing the list with the control stop word lists. The technique which consistently outperformed the other techniques will be considered to be the most effective stop word list generator.

## Results and Analysis

Using the experimental approach described above, we have generated the various stop words lists and compared their performance at detecting stop words which are listed in the Fox's and Manu's list. In the following sections, we will first compare the various measures and their performance with the Manu's list which is the smaller of the two lists. After the initial comparison, we will then further compare the results using the Fox's list for a second level of validation. The results are plotted with the F-Measures and the threshold levels.

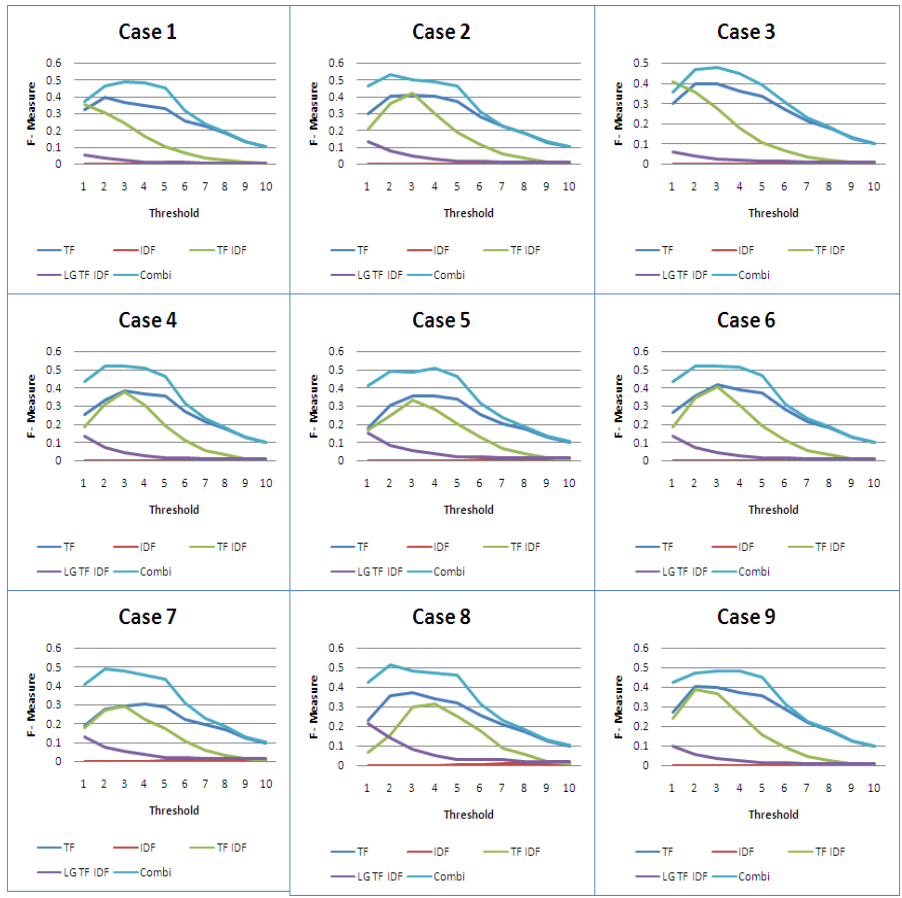

Chart 1: Comparison of the performance of the measures with the Manu's List

From the chart 1, we can see that the combination technique outperforms most of the other techniques by a fair margin. With the exception of a few initial threshold, where TF*IDF or Log(TF)*IDF variant performs better, the new proposed approach is distinctly better than the other techniques. This superior performance could be attributed to the smaller list of stop words generated by combination approach compared to the other techniques. This effect is further compounded by the small list of stop words in the Manu instance. Many of the words included in the new stop word lists include new words which could be stop words in the context of the twitter contents.

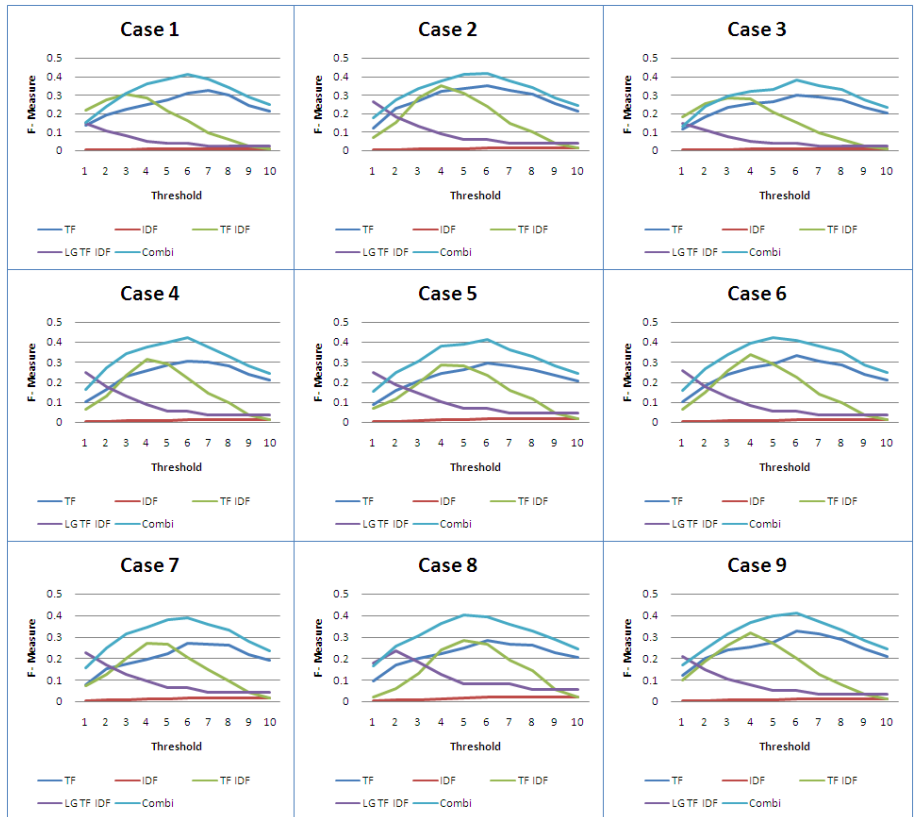

Chart 2: Comparison of the performance of the measures with the Fox's List

From the chart 2, we can see that the combination technique outperforms most of the other techniques by a fair margin. However, the technique is not as strong as some of the other techniques in the initial threshold levels in some cases. The drop in performance could be attributed to the larger list of stop words covered by Fox's list which is almost three times the size of Manu's list. At the same time, as mentioned earlier, the stop word list generated by the combination technique is also smaller than its TF*IDF and variant counterparts. However, the combination technique still outperforms the other techniques beyond the initial threshold which indicates its superior performance on the overall.

**Conclusion**

In this paper, we proposed a new method for automatically generating a stop word list for a given collection of tweets. The approach is based on the combinatorial nature of the words in speeches. We investigated the effectiveness and robustness of the approach by testing it against 9 collections of tweets from different periods. The approach is also compared with the existing approaches using TD*IDF and variants. The results indicated that the new approach is comparable to existing approaches if not better in certain cases.

The direct nature of the combinatorial approach is not normalized and additional research is needed to produce the normalized measure. Other newer approaches such as page-rank approach will also require more research to understand the effectiveness. Future research will also need to investigate the scenario of three or more combinations of words to determine whether they are stop words.